\documentclass[manuscript]{emulateapj}
\usepackage{graphicx}
\usepackage{subfigure}
\usepackage{natbib}
\usepackage{amsmath}
\usepackage{color}
\usepackage{times}

\newcommand\teff{$T_{\rm eff}$}
\newcommand\logg{log\,{\it g}}

\newcommand\etal{\rm et al.\,}

\newcommand\SDSS{SDSS J102915+172927}
\newcommand\SDSSx{SDSS J1029+1729}
\newcommand\SMSS{SMSS J031300.36-670839.3}
\newcommand\SMSSx{SMSS\,0313-6708}
\newcommand\BD{BD+39~4926}

\slugcomment{Accepted for publication in The Astrophysical Journal}
\shorttitle{Dust around Metal-Poor Stars} 
\shortauthors{Venn \etal}
\begin{document}

\title{Searching for Dust around Hyper Metal-Poor Stars\altaffilmark{$\dagger$}}

\altaffiltext{$\dagger$}{Based on observations obtained at the Gemini Observatory, which is operated by the Association of Universities for Research in Astronomy, Inc., under a cooperative agreement with the NSF on behalf of the Gemini partnership: the National Science Foundation (United States), the Science and Technology Facilities Council (United Kingdom), the National Research Council (Canada), CONICYT (Chile), the Australian Research Council (Australia), Minist\'{e}rio da Ci\^{e}ncia, Tecnologia e Inova\c{c}\~{a}o (Brazil) and Ministerio de Ciencia, Tecnolog\'{i}a e Innovaci\'{o}n Productiva (Argentina).}

\author{Kim A. Venn$^{1}$, 
        Thomas H. Puzia$^{2}$,
        Mike Divell$^{1}$, 
        Stephanie C{\^o}t{\'e}$^{3}$,
        David L. Lambert$^{4}$, 
        Else Starkenburg$^{1}$ 
}

\affil{$^{1}$Department of Physics \& Astronomy, University of Victoria, 
             3800 Finnerty Road, 
             Victoria, BC, V8P 5C2, Canada \\
  $^{2}$Institute of Astrophysics, 
        Pontificia Universidad Catolica de Chile,
        Av.~Vicu\~na Mackenna 4860, 7820436 Macul, Santiago, Chile\\
  $^{3}$NRC Herzberg Institute of Astrophysics, 5071 West Saanich Road,
             Victoria, BC, V9E 2E7, Canada\\
  $^{4}$McDonald Observatory and the Department of Astronomy, 
        University of Texas at Austin, RLM 15.308,
             Austin, TX, 78712, USA
}
\email{kvenn@uvic.ca}

\begin{abstract}
We examine the mid-infrared fluxes and spectral energy distributions for metal-poor stars with iron abundances [Fe/H]~$\lesssim-5$, as well as two CEMP-no stars, to eliminate the possibility that their low metallicities are related to the depletion of elements onto dust grains in the formation of a debris disk.~Six out of seven stars examined here show no mid-IR excess.~These non-detections rule out many types of circumstellar disks, e.g.~a warm debris disk ($T\!\le\!290$\,K), or debris disks with inner radii $\le$1\,AU, such as those associated with the chemically peculiar post-AGB spectroscopic binaries and RV Tau variables.~However, we cannot rule out cooler debris disks, nor those with lower flux ratios to their host stars due to, e.g.~a smaller disk mass, a larger inner disk radius, an absence of small grains, or even a multicomponent structure, as often found with the chemically peculiar Lambda Bootis stars.~The only exception is HE0107-5240, for which a small mid-IR excess near 10 microns is detected at the 2-$\sigma$ level; if the excess is real  and associated with this star, it may indicate the presence of (recent) dust-gas winnowing or a binary system.  
\end{abstract}

\keywords{stars: abundances, chemically peculiar, individual stars (HE0107-5240),
metal-poor stars; debris disks}

\section{Introduction \label{intro}}

It is well established that only light elements were synthesized in the Big Bang, leaving heavier nuclei from Carbon to Uranium to be formed later in stars \citep[e.g.,][]{coc14, coc12, rya00}.~Thus, a fossil record of the earliest episodes of stellar nucleosynthesis in the local universe is likely to be revealed by the compositions of the most metal poor Galactic stars \citep[e.g.][]{ume03, tum07a, tum07b, tom07}.~While it was originally speculated that the first generation of stars consisted of massive stars only \citep[${\cal M}_\star\!>\!100\,M_\odot$, e.g.][]{ost96, ume03}, later simulations that account for fragmentation are able to form low-mass stars even without metals through molecular hydrogen line cooling \citep[e.g.][]{nak01, tur09, cla11, sch03, sch12, bro13}. \cite{gre11} suggest that the IMF of the first stellar generation may have reached only a maximum stellar mass of $10\,M_\odot$ due to fragmentation, but also that this IMF would be top-heavy.~Stars with masses of $\leq0.8\,M_\odot$ would have main sequence lifetimes that exceed the age of the universe \citep[$13.8\!\pm\!0.06$\,Gyr,][]{pla03}, and therefore should still be observable today.~Since stellar evolution theory predicts only minor changes in the chemical composition of their stellar atmospheres over their main-sequence lifetimes \citep[e.g.~due to atomic diffusion, convection, and rotation that alter very few elemental abundances, except possibly Lithium, see][]{cas96, sch03, ven02, pic04, mey06, eks12, gru13}, these stars are expected to hold the memory of the chemical composition of the primordial gas from which they have formed.    
 
The lure of this revelation has driven the search to find and analyze such Rosetta stones.~A great leap forward was achieved with the discovery of two stars with iron abundances [Fe/H]~$< -5.3$, the red giant HE0107-5240 \citep{chr02, chr04, lim03, sud04} and the subgiant HE1327-2326 \citep{fre05, aok06, kor09, bon12b} -- defining the `hyper metal poor' (HMP) stars \citep{bee05}.~Two more stars with [Fe/H] $\simeq\!-5$ have been discovered since then; HE0557-4840 \citep{nor07, nor12} and \SDSS\ (hereafter \SDSSx) \citep{caf12, bon12a}.~The latter star is particularly interesting because it shows no enhancements in CNO, unlike the other HMP stars, and since these three elements are amongst the most abundant by number in any star, including the Sun \citep{asp09}, then the total number of atoms in the latter star are in fact lower than the others, making it the most metal-poor star yet known.~More recently, an even more iron-poor star, \SMSS\ (hereafter \SMSSx), has been found \citep{kel14}, with [Fe/H] $< -7$ but also C-rich.  At such low metallicities, there are very few spectral absorption lines, and abundance measurements for only four elements (Li, Ca, Mg, and C) have been published.

\begin{center}
\begin{deluxetable*}{ccllllll}
\footnotesize
\tablecaption{Magnitudes and Flux Densities for the Metal Poor Stars\label{tab:mags}}  
\tablewidth{0pt}
\tablehead{
\colhead{Band} & \colhead {$\lambda_c$} & \colhead{HE0107-5240} &  \colhead{HE0557-4840} & \colhead{HE1327-2326} & \colhead{\SDSSx}  & \colhead{\SMSSx} & \colhead {Refs} \\
\colhead{} & \colhead {(nm)} & \colhead{mag \,\, flux} &  \colhead{mag \,\, flux} & \colhead{mag \,\, flux} & \colhead{mag \,\, flux}  & \colhead{mag \,\, flux} & \colhead {} 
} 
\startdata
U & 373  & \nodata & \nodata & 13.77 \,\, 194.40 & \nodata & \nodata & (1) \\
B & 444  &  15.84 \,\, 31.01 & 16.17 \,\, 21.87 & 13.97 \,\, 223.30 & \nodata & \nodata & (1,2) \\
V & 548  &  15.17 \,\, 32.61 & 15.45 \,\, 24.16 & 13.53 \,\, 178.50 & 16.73 \,\, 7.46 & 14.7 \,\, 48.38 & (1,2,3,4) \\
R & 686  &  14.73 \,\, 25.43 & 14.99 \,\, 19.36 & 13.16 \,\, 125.50 & \nodata & \nodata & (1,2) \\
I & 864  &  14.28 \,\, 18.53 & \nodata     & 12.80 \,\, 79.45  & \nodata & \nodata & (1,2) \\
\\
J & 1235 & 13.676 \,\, 10.59 & 13.792 \,\, 9.52 & 12.357 \,\, 35.69 & 15.51 \,\, 1.96 & 13.181 \,\, 16.71 & (5) \\
H & 1662 &  13.253 \,\, 5.66 & 13.304 \,\, 5.40 & 12.068 \,\, 16.87 & 15.13 \,\, 1.01 & 12.692 \,\, 9.49 & (5) \\
K & 2159 &  13.218 \,\, 2.21 & 13.194 \,\, 2.26 & 11.986 \,\, 6.88  & 15.15 \,\, 0.37 & 12.661 \,\, 3.69 & (5) \\
\\
N & 10360 & $<$12.66 $<$1.05 & $<$13.21 $<$0.64 & \nodata & \nodata & \nodata & (6) \\
\\
W$_1$ & 3350  &   13.088 \,\, 47.63 & 13.150 \,\, 44.99 &  11.961 \,\, 134.50 &    14.908 \,\, 8.91 & 12.534 \,\, 79.35 & (7) \\
W$_2$ & 4600  &   13.105 \,\, 13.83 & 13.169 \,\, 13.04 &  11.957 \,\, 39.82  &    14.681 \,\, 3.24 & 12.564 \,\, 22.77 & (7) \\
W$_3$ & 11560 &   12.285 \,\, 0.79  & 13.224 \,\, 0.33  &  12.301 \,\, 0.78   & $<$12.665 $<$0.56   & 12.684 \,\, 0.55 & (7) \\
W$_4$ & 22080 & $<$8.648 $<$1.77 & $<$9.711 $<$0.66 & $<$8.718 $<$1.66 & $<$9.030 $<$1.24 & $<$9.710 $<$0.66 & (7) \\
\\
A$_{\rm V}$ & \nodata & 0.04 & 0.13 & 0.22 & \nodata & \nodata & (1,2,3,4,8) \\\vspace{-0.2cm}
\enddata
\tablecomments{
Fluxes x 10$^{-16}$ erg s$^{-1}$ cm$^{-2}$ A$^{-1}$, except for the mid-IR WISE data 
and Gemini TReCs $N$-band data, which are Fluxes x 10$^{-18}$ erg s$^{-1}$ cm$^{-2}$ A$^{-1}$.
(1) \cite{bee07},
(2) \cite{chr02},
(3) \cite{caf12},
(4) \cite{kel14},
(5) 2MASS Point Source Catalogue \citep[]{skr06},
(6) Gemini TReCS observations (this paper),
(7) WISE all sky survey \citep[]{wri10},
(8) Interstellar extinction A$_{\rm V}$ = 3.1 $E(B-V)$, or set to zero when reddening is unavailable.   
}
\end{deluxetable*}
\end{center}

\begin{deluxetable}{lcccl}
\footnotesize
\tablecaption{Stellar Parameters for Model Atmospheres\label{tab:atms}}  
\tablewidth{0pt}
\tablehead{
\colhead{Star} & \colhead{\teff} & \colhead{\logg} &  \colhead{[Fe/H]} & 
\colhead{Refs}
} 
\startdata
HE1327-2326 & 6250 & 3.5 & $-$5.0 & (1) \\
HE0557-4840 & 5000 & 2.0 & $-$5.0 &  (2) \\
HE0107-5240 & 5000 & 2.0 & $-$5.0 &  (3) \\
\SDSSx      & 5500 & 4.0 & $-$5.0 &  (4) \\
\nodata     & 5750 & 4.0 & $-$5.0 &  (4) \\
\SMSSx      & 5000 & 2.5 & $-$5.0 &  (5) \\
\nodata     & 5250 & 2.5 & $-$5.0 &  (5) \\
\\
Comparisons: \\
CS22949-037 & 5000 & 1.5 & $-$4.0 &  (6) \\
BD+44~493   & 5500 & 3.5 & $-$4.0 &  (7) \\
BD+39~4926  & 7750 & 2.5 & $-$2.5 &  (8) \\
HP~Lyr      & 6250 & 1.0 & $-$1.0 &  (9) \\ \vspace{-0.2cm}
\enddata
\tablecomments{
The parameters for the online MARCS (and some ATLAS) model atmospheres
adopted are listed, which are closest to the stellar parameters 
determined in the following references;
(1) \cite{fre05},
(2) \cite{nor07},
(3) \cite{chr02},
(4) \cite{caf12},
(5) \cite{kel14},
(6) \cite{spi11},
(7) \cite{ito13},
(8) \cite{rao12},
(9) \cite{gir05}.
}
\end{deluxetable}

Prior to these remarkable discoveries, the most Fe-poor stars known were HR\,4049 and HD\,52961, both with [Fe/H]~$<\!-4.8$ \citep[e.g.][]{wae91}.~However, these and slightly more iron-rich examples were dismissed -- correctly -- as irrelevant to the issue of early stellar nucleosynthesis because they are chemically peculiar, post-AGB, spectroscopic binaries, i.e.~their present surface compositions are far removed from their initial values.~Initial compositions of such stars are revealed by those observable elements least likely to condense out into grains: C, N, O, S and Zn, with the additional recognition that C, N, and O are very probably altered by internal mixing during the course of stellar evolution.~For example, \cite{tak02} reported that HR\,4049 has [Fe/H] = $-4.7$, but [S/H] = $-0.5$ and [Zn/H] = $-1.3$.~These latter abundances suggested an initial composition closer to [Fe/H] = $-1.0$ for which unevolved, normal stars have [S/Fe] = +0.4 and [Zn/Fe] = 0.

Thus, the current compositions of these chemically peculiar stars reflect that of gas from which refractory elements have been removed to varying degrees by the process of `dust-gas winnowing' \citep{hin07}.~In gas of sufficiently low temperature, dust condenses out and the gas is depleted in those elements that form the dust. The local interstellar gas, for example, displays such depletions \citep{sav96}.~If a star were to accrete gas (or re-accrete gas, if the separation occurs in a circumstellar disk/shell), preferentially over dust, and the accreted gas were to comprise a major fraction of the stellar atmosphere, then the star will exhibit striking abundance anomalies.

Anomalies plausibly attributed to dust-gas winnowing have been reported for post-AGB stars (spectroscopic binaries and RV Tauri variables), and the class of main sequence A-type known as Lambda Bootis stars.~These stars occupy significantly different parts of the HR-diagram and, therefore, associated properties of affected stars, such as age, mass, binarity, or evolutionary status, do not appear to be uniform across the cases. Details in the observed elemental abundance patterns vary between the cases as well, possibly due to differences in the stellar parameters (e.g.~temperature, gravity), the stellar atmospheres (e.g.~convection, mixing), or perhaps due to the dust-gas winnowing mechanism itself.~The principal common features linking these post-AGB and main sequence stars are the presence of dust and a thin outer envelope.

\begin{figure*}[!t]
\centering
\includegraphics[width=15.8cm]{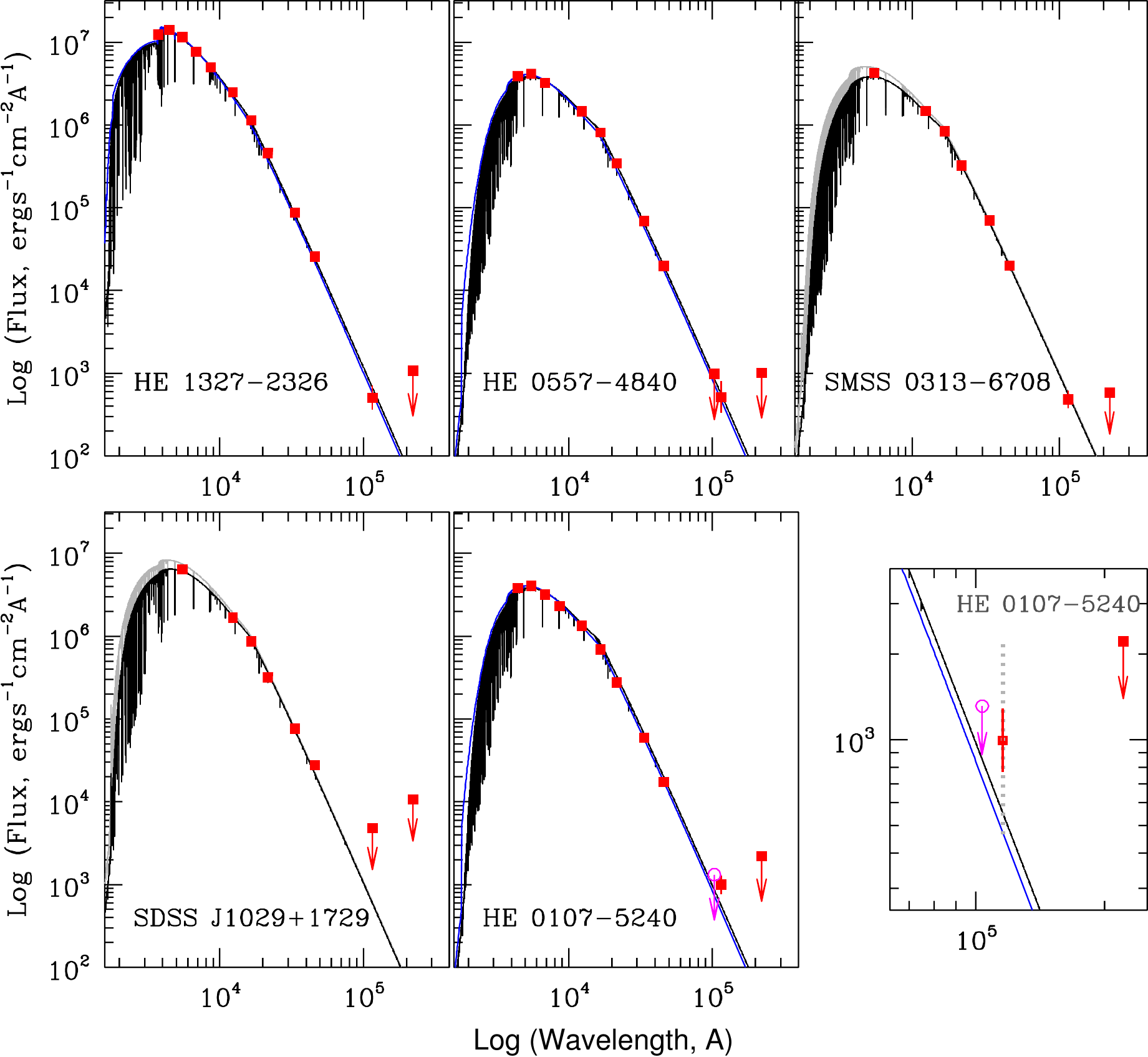}
\caption{The spectral energy distribution for the HMP and other stars; HE1327-2326 (Frebel \etal 2005; top left), HE0557-4840 (Norris \etal 2007; top middle) \SMSSx\, (Keller \etal 2014; top right), \SDSSx, (Caffau \etal 2012; bottom left), HE0107-5240 (Christlieb \etal 2002; bottom middle, with bottom right zoomed inset of the mid-IR range).~Broadband photometric fluxes are listed in Table~\ref{tab:mags} (red squares), and $N$-band upper-limits determined from Gemini TReCs data (pink squares; see Appendix).~The model atmosphere SEDs are from MARCS (black curve) and Kurucz models (blue curve), with atmospheric parameters listed in Table~\ref{tab:atms} for each star.~When two MARCS model fluxes are compared, the hotter model is noted with a grey curve.~Observed fluxes are scaled to the model fluxes in order to best match the $R, I$, and $J$-band magnitudes.~The 1-$\sigma$ photometric uncertainties are noted when larger than the symbol size, except in the case of the inset for HE0107-5240 where the 3-$\sigma$ uncertainy is also shown by a grey dotted line.}
\label{fig:sedbig}
\end{figure*}

A comparison of the HMP stars HE1327-2326 and HE0107-5240, as well as HE0557-4840, to samples of these chemically peculiar stars led \cite{ven08} to speculate on the question of whether the HMP stars may themselves be chemically peculiar.~While \cite{ven08} left the question unresolved, \cite{chr04}, who also considered the possibility of selective dust depletion for E0107-5240, sided against the possibility due to a high C/N ratio and unknown pulsation or binary properties typically seen in the chemically peculiar stars; however, they also suggested that the Zn and S abundances should be investigated as volatile elements that do not easily condense onto dust grains.~This test was initiated for HE1327-2326, when \cite{bon12b} attempted to determine its sulphur abundance -- unfortunately, the upper limit determined for Sulphur was inconclusive regarding the likelihood of dust formation affecting the surficial chemical abundance ratios.~\cite{bon12b} also noted that dust formation and supernova yields with extensive fall-back are not mutually exclusive (nor is an alteration to C/N due to convective mixing) such that the abundance ratios in the HMP stars could be affected by more than one mechanism.   

If the HMP stars are, in fact, chemically peculiar stars, i.e.~normal stars that have had their chemical abundances altered by dust-gas winnowing, then their initial abundances would have been higher ([Fe/H] $\sim$ $-2$ to $-4$, based on their CNO and Na abundances).~This would make them typical metal poor stars, although ones that have undergone a rare process.~The dust removed/not accreted in their atmospheres should then be present in a disk or shell outside the stellar envelope.~Therefore, another test of the dust-gas separation hypothesis is to search for the presence of circumstellar dust, e.g.~to search for an infrared excess in these stars.~No near-IR excess through the $K$ band (2.2 microns) is seen in the colors of HE1327-2326, HE0107-5240, nor HE0557-4840 compared to model flux distributions \citep{ven08}.~However, the presence of circumstellar material can be difficult to detect and is often inconclusive from $JHK$ magnitudes alone, e.g.~some post-AGB and chemically peculiar Lambda Bootis stars only show IR excesses beyond 2.2 microns as discussed below.

To further explore the possibility that the HMP stars have been affected by dust-gas winnowing, we examine these stars and other rare metal-poor stars at wavelengths redward of the classic near-IR passbands $JHK$, to search for circumstellar material.~In this paper, we present the first examination of the mid-infrared fluxes of the HMP and other metal-poor stars from Gemini-South TReCs imaging (GS-2009B-Q88) and the WISE all sky survey\footnote{The Wide-field Infrared Survey Explorer All-Sky Data Release from March 14, 2012 at "http://wise2.ipac.caltech.edu/docs/release/allsky/".}.

\begin{figure*}[t]
\centering
\includegraphics[width=15.75cm]{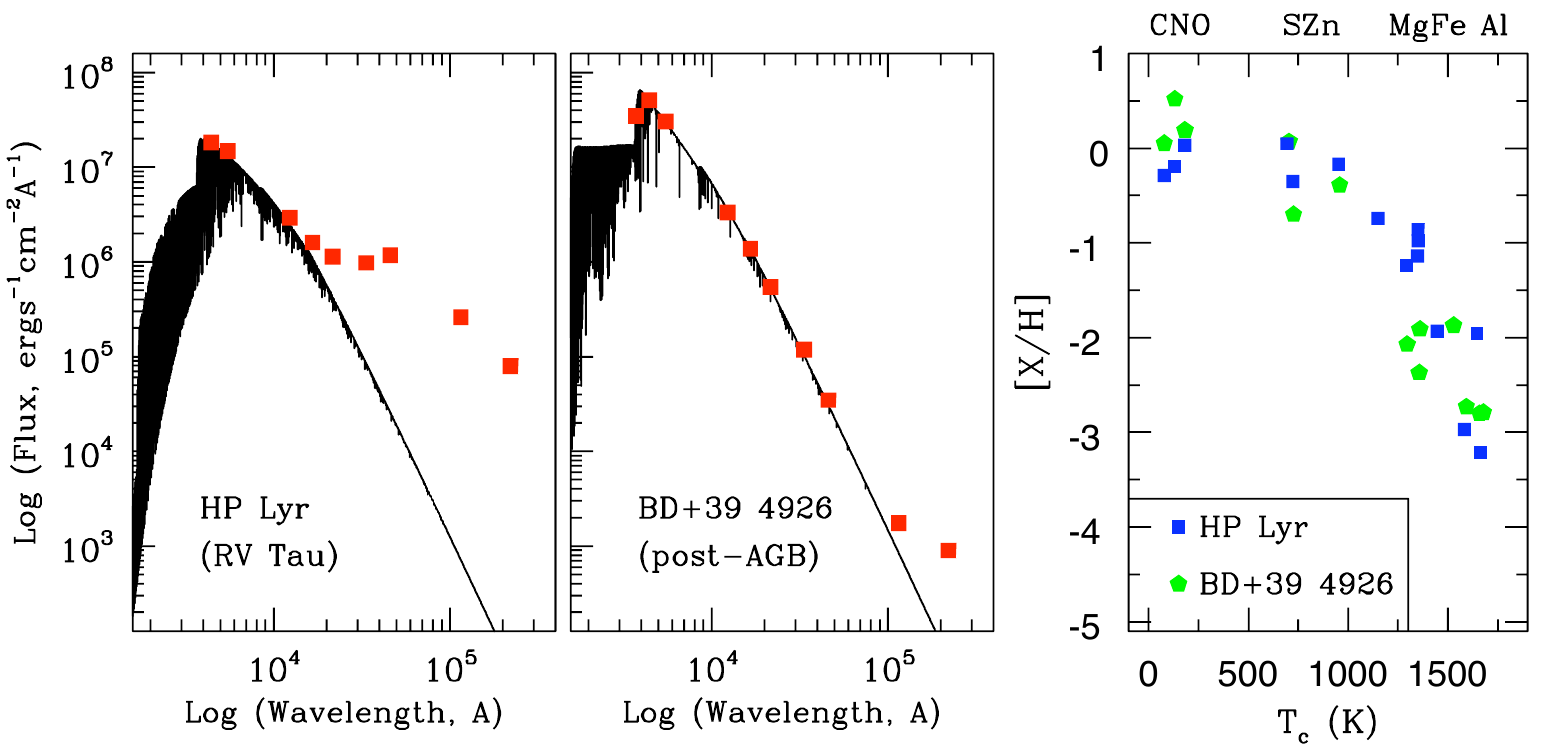}
\caption{The spectral energy distribution for two chemically peculiar stars that show mid-IR excesses; the RV~Tau variable HP~Lyr (left panel) and the post-AGB star \BD\, (middle panel).~MARCS models with the atmospheric parameters from \cite{rao12} and \cite{gir05} are shown.~Broadband $BVR$ photometric colors from \cite{fuj02}, $JHK_s$ from the 2MASS Point Source Catalogue, and mid-IR magnitudes from the WISE All-Sky Source Catalogue.~A reddening estimate of $E(B\!-\!V)\!=\!0.2$ was adopted for HP Lyr.~Observations are scaled to absolute flux units, and the photometric uncertainties are smaller than the symbol sizes.~Chemical abundances for these stars from \cite{rao12} and \cite{gir05} versus the element condensation temperature \citep[based on][]{lod03} are shown in the right panel.}
\label{fig:sed4}
\end{figure*}

\section {Spectral Energy Distributions } 

\subsection {Hyper Metal-Poor Stars, and Others} 

Photometric magnitudes for the HMP and other metal-poor stars have been taken from the literature; see Table~\ref{tab:mags}.~$BVRI$ magnitudes have been converted to fluxes using the Vega zero point calibrations by \cite{col96}; $JHK$ magnitudes were converted using values in the 2MASS Explanatory Supplement (Section VI.4a).~The WISE fluxes were converted using the WISE All-Sky Release Explanatory Supplement (Section IV.4.h.v).~$N$-band magnitude limits are those determined from Gemini TReCS imaging as described in the Appendix.~Reddening corrections have been applied when available, but $E(B\!-\!V)$ is very small for the metal-poor (halo) stars, and has no significant effects, especially in the IR.

The spectral energy distributions (SEDs) for the HMP and other metal-poor stars are shown in Figure~\ref{fig:sedbig}.~The observed fluxes are compared to model flux distributions from the online grid of MARCS \citep{gus75, asp97, gus08} and Kurucz models \citep{cas04}.~The model parameters were adopted from the literature; see Table~\ref{tab:atms}.~There are very small differences in the SEDs between the MARCS and Kurucz model atmospheres, mainly due to differences in the opacities and the variation in the MARCS models as a function of stellar mass and [$\alpha$/Fe] ratio.~However, these variations are insignificant for our purposes.
 
The observed fluxes were scaled to match the model fluxes best in the $R$, $I$ and $J$ filters; the scale factor is equivalent to the ratio of (distance/radius)$^2$.~For \SMSSx\, and \SDSSx, MARCS model SEDs are shown for two different temperatures to account for uncertainties in their determinations, but the differences in the SEDs are negligible in the IR.~No reddening corrections are assumed for \SMSSx, which may affect the $V$-band flux determination but has no significant impact on the IR fluxes.

Inspection of Figure~\ref{fig:sedbig} shows that with one possible exception the long-wavelength tail of the SEDs out to about 10 microns is dominated by the contribution from the star and a dust signature is not detected.~The possible exception is HE0107-5240, where the WISE W3 flux at 11.6 microns is marginally greater than that predicted for this star; a close-up view of its mid-IR flux shows the excess is no more than 2 $\sigma$ of the error in the W3 flux for this star as reported in the WISE All sky survey, and warrants deeper observations.~If thermal emission from a dusty disk peaks at this wavelength, the implied temperature would be $T\!\le\!300$\,K.~Further inspection of the WISE W2 4.6 micron flux for \SDSSx, is also intriguing, possibly suggesting a slight upturn from the predicted flux for this star; however, the WISE W3 and W4 magnitudes provide only upper limits that do not further constrain its fluxes.~Small (1 $\sigma$) differences in the model atmosphere parameters have no significant effect on the overall comparisons.  

\begin{figure*}[t]
\centering
\includegraphics[width=15.75cm]{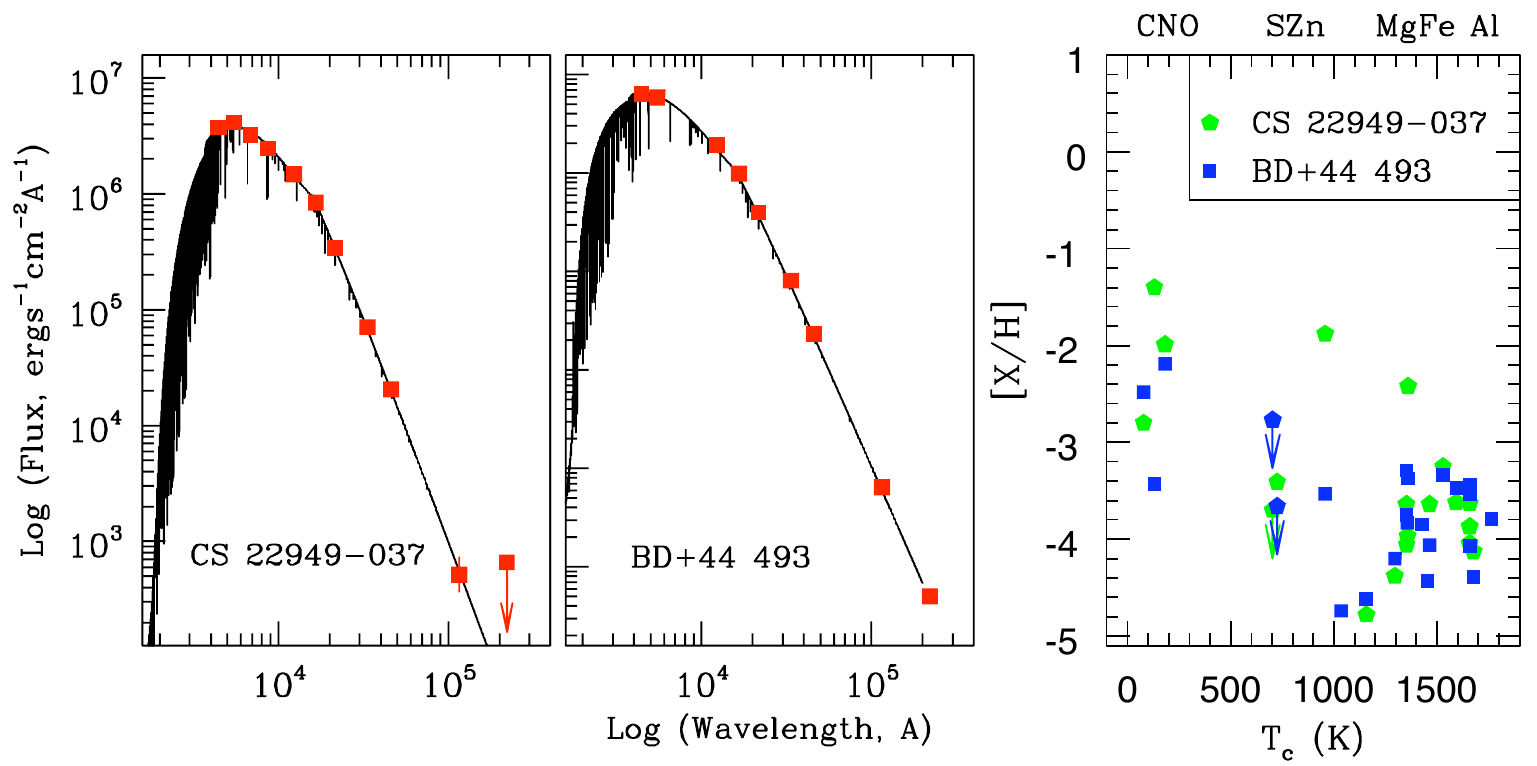}
\caption{The spectral energy distribution for two CEMP-no stars, BD+44~493 (left panel) and CS22949-037 (middle panel).~MARCS models with the atmospheric parameters from \cite{spi11} and \cite{ito13} are shown.~Broadband $BVRI$ photometric colors from \cite{bee07} and \cite{hog00}, $JHK_s$ from the 2MASS Point Source Catalogue, and mid-IR magnitudes from the WISE All-Sky Source Catalogue.~Observations are scaled to the absolute flux units, and the photometric uncertainties are smaller than the symbol sizes, with the exception of WISE W3 for CS22949-037.~Both stars show no evidence for mid-IR excesses, particularly the brighter star BD+44~493 which shows no excess at 22 microns.~These are the only two CEMP-no stars with known low [S/H] and/or [Zn/H] abundances (see right panel).}
\label{fig:sed1}
\end{figure*}

\subsection{Chemically Peculiar Post-AGB Stars and RV Tau Variables}

Post-AGB spectroscopic binaries and RV Tauri variable stars are two kinds of chemically peculiar stars with features that are plausibly attributed to dust-gas winnowing.~Post-AGB stars can show strong IR emission in the $K$-band \citep[like HR~4049,][]{dom03}, though most do not show an IR excess until longer wavelengths, e.g.~beyond 10 microns (the $N$ band).~This can be seen in the figures for C-rich and O-rich post-AGB stars from \cite{van03}; and we show two examples in Figure~\ref{fig:sed4}, the bright RV Tau variable HP~Lyr, and the post-AGB binary star \BD.~Photometric data for these two stars is collected from the literature (see Table~\ref{tab:comps}), and converted to fluxes using the same methods as above for the HMP stars.~Stellar parameters are those determined by \cite{rao12} and \cite{gir05}.~As above, the observed fluxes were scaled to match the model fluxes best in the $R$, $I$ and $J$ filters.

To illustrate the signature of dust-gas winnowing in these two stars, we also show their stellar chemical abundances compared to the condensation temperature of various elements in Figure~\ref{fig:sed4}.~Abundances are determined from detailed LTE model atmosphere analyses of high resolution, high signal-to-noise spectra by \cite{rao12} and \cite{gir05}.

It is worth noting that \BD\, had been thought to be free of IR emission \citep{rao12}, which is unusual given the strong chemical anomalies that are consistent with the dust-gas winnowing signature.~However, the new WISE data presented here reach deeper magnitudes and reveal the mid-IR excess that clearly accompanies this star beyond 10 microns.~{\it This further implies that the IR excess of a star that has undergone dust-gas winnowing does not correlate with the strength of the chemical anomalies}.~The formation, evolution, and detection of a debris disk is a delicate balance between temperature and radius, mass and density, composition and emissivity, and especially age of formation and time for dispersal -- all relative to the conditions in the stellar atmosphere, where the chemical anomalies form, but can also be erased through convective mixing with the stellar interior.

\subsection{CEMP-no Stars}

Four out of five of the stars with [Fe/H] $\lesssim -5$ are carbon-enhanced, thus CEMP stars (carbon-enhanced metal poor stars).~The CEMP stars do not comprise a single stellar population or formation mechanism.~They are known to show distinct chemical and binary signatures that separate them into at least two separate classes: CEMP-s, CEMP-no, and possibly a third class, the CEMP-r/s stars.~The CEMP-s stars are nearly certainly the low metallicity end of CH-stars, which are enriched through binary mass transfer with an AGB companion (Lucatello \etal 2005, McClure \etal 1997).~The CEMP-no stars show low neutron capture ratios unlike the CEMP-s stars (defined by [Ba/Fe]~$\!<\!0$), and the complete CEMP-no dataset is clearly inconsistent with the binary properties of the CEMP-s class (Aoki \etal 2010, Norris \etal 2013, Starkenburg et al. 2014).~The CEMP-r/s stars are also associated with mass transfer in a binary system, although the stars may have been previously enriched in r-process elements (Allen et al. 2012, Bisterzo \etal 2010, Herwig \etal 2014). 
 
\cite{spi11} point out that one CEMP-no star, CS\,22949-037, with [Fe/H]~$\!=\!-4.0$, has low [S/Fe] and [Zn/Fe] abundance ratios.~These two elements are critical in testing for dust formation because they have low condensation temperatures and do not deplete onto dust grains easily.~Without an excess of S and Zn, this star does not appear to be affected by dust-gas winnowing, should not show a mid-IR excess from any debris disk, and therefore should have retained its natal chemical composition.~BD+44\,493 (Ito \etal 2013) has a low [Zn/Fe] upper limit which also suggests no significant dust depletions.~The abundance pattern for these two stars is shown in Figure~\ref{fig:sed1}, and the low Zn and S abundances (as well as overall dispersion in element ratios) stand out as distinct from the chemical abundance pattern of the chemically peculiar stars shown in Figure~\ref{fig:sed4}.

The SEDs for these two CEMP-no stars are shown in Figure~\ref{fig:sed1}.~Photometric colours are taken from the literature (see Table~\ref{tab:comps}) and converted to fluxes using the same zero point calibrations cited above for the HMP stars.~The observed fluxes are compared to model flux distributions from the online grid of MARCS models \citep{gus75, asp97, gus08}, with the stellar parameters determined by \cite{spi11} and \cite{ito13}.~As above, the observed fluxes were scaled to match the model fluxes best in the $R$, $I$ and $J$ filters.~The comparison of the data to the model fluxes are exquisite.~In the case of the brighter star, BD+44~493, the fit extends to 22 microns, the WISE W4 magnitude, where the 1 $\sigma$ uncertainty in the W4 flux is smaller than the size of the data point on the figure.~There is absolutely no evidence for IR excesses around these two CEMP-no stars.
 
\begin{center}
\begin{deluxetable*}{cclllll}
\footnotesize
\tablecaption{Magnitudes and Flux Densities for Chemically Peculiar Comparison Stars\label{tab:comps}}  
\tablewidth{0pt}
\tablehead{
\colhead{Band} & \colhead {$\lambda_c$} & \colhead{HP~Lyr} &  \colhead{\BD} & \colhead{BD+44~493} & \colhead{CS\,22949-037}  & \colhead {Refs} \\
\colhead{} & \colhead {(nm)} & \colhead{mag \,\, flux} &  \colhead{mag \,\, flux} & \colhead{mag \,\, flux} & \colhead{mag \,\, flux}  & \colhead {} 
} 
\startdata
U & 373  & \nodata   & 9.61 \,\,  126.1  & \nodata     & \nodata & 2 \\
B & 444  & 11.02 \,\, 54.78  & 9.49 \,\,  184.3  & 9.63 \,\,  89.99 & 15.12 \,\, 0.69 & 1,3 \\
V & 548  & 10.43 \,\, 44.53  & 9.29 \,\,  109.8  & 9.11 \,\,  83.30 & 14.36 \,\, 0.76 & 1,3 \\
R & 686  & \nodata   & \nodata      & \nodata     & 13.90 \,\, 0.59 & 1 \\
I & 864  & \nodata   & \nodata      & \nodata     & 13.36 \,\, 0.45 & 1 \\
\\
J & 1235 &  8.877 \,\, 8.802  & 8.539  \,\, 12.02 & 7.659 \,\,  27.03  & 12.650 \,\, 0.27 & 4 \\
H & 1662 &  8.429 \,\, 4.815  & 8.394  \,\,  4.97 & 7.269 \,\,  14.02  & 12.159 \,\, 0.15 & 4 \\
K & 2159 &  7.750 \,\, 3.402  & 8.338  \,\,  1.98 & 7.202 \,\,   5.64  & 12.075 \,\, 0.06 & 4 \\
\\
W$_1$ & 3350  & 6.118 \,\, 292.4 & 8.199 \,\, 43.01 & 7.134 \,\, 114.70  & 11.985 \,\, 1.316 & 5 \\
W$_2$ & 4600  & 4.585 \,\, 353.9 & 8.207 \,\, 12.59 & 7.165 \,\,  32.88  & 12.005 \,\, 0.381 & 5 \\
W$_3$ & 11560 & 2.310 \,\, 77.61 & 7.528 \,\,  0.63 & 7.131 \,\,  0.915  & 12.082 \,\, 0.009 & 5 \\
W$_4$ & 22080 & $<$0.825 $<$23.81 & 5.482 \,\, 0.33 & 7.133 \,\,  0.071  & $<$9.050 $<$0.012 & 5 \\
\\
A$_{\rm V}$ &  & 0.62 & 0.47 & \nodata & 0.18 & 1,2,3,6 \\\vspace{-0.2cm}
\enddata
\tablecomments{
Fluxes x 10$^{-14}$ erg s$^{-1}$ cm$^{-2}$ A$^{-1}$, except for the mid-IR WISE data 
which are Fluxes x 10$^{-16}$ erg s$^{-1}$ cm$^{-2}$ A$^{-1}$.
(1) \cite{bee07},
(2) \cite{kle62},
(3) \cite{hog00},
(4) 2MASS Point Source Catalogue \citep[]{skr06},
(5) WISE all sky survey \citep[]{wri10},
(6) Interstellar extinction A$_{\rm V}=3.1\times E(B-V)$, or set to zero when reddening is
    unavailable.~For HP~Lyr, $E(B-V) = 0.2$ is assumed, to improve the B and V flux comparisons.
}
\end{deluxetable*}
\end{center}

\section{Discussion}

The main questions in this paper are whether the HMP stars have debris disks, and whether the formation of a debris disk has affected the chemical abundances in their stellar atmospheres.~Ascertaining the integrity of the HMP stars as old and intrinsically metal poor is important since these stars are used as benchmarks for the earliest stages of chemical evolution in the Galaxy, and therefore as constraints for yields from the earliest supernova.

\subsection{Formation and Detection of Debris Disks} 

Detection of debris disks is challenging, although the new mid-IR ground-based (ALMA) and space-based (Herschel, WISE) telescopes have made significant improvements.~Emission from a debris disk depends on two parameters in a simple one zone model -- dust temperature and fractional luminosity \citep{wya08}.~Fractional luminosity depends on the ratio of fluxes, disk vs.~star at their respective blackbody peak wavelengths, thus, f = (F$_{{\nu} {\rm disk, max}}$ / F$_{{\nu} {\rm *, max}}$)/($\lambda_{\rm *, max}$/$\lambda_{\rm disk, max}$).~If the disk flux density were to peak at 10 microns, the lack of a detection in the $N$ band or WISE W3 band will limit the flux ratio between debris disk and host star to f $\le 10^{-4}$.~This also limits the distance to the inner radius of a potential debris disk from a host star (with $T_{\rm eff}\!=\! 5000$\,K) to $\ge\!0.7$ AU.~These numbers are similar to those for close-in A-star disks \citep[]{wya08}, e.g., $\zeta$ Lep hosts a dusty disk with $T\!\approx\!320$\,K that is detected only beyond 2\,AU \citep{moe07}.

Observations taken with the Herschel telescope of debris disks have been carried out for A-type stars \citep[the DEBRIS survey]{boo13} and FGK stars \citep[the DUNES survey]{eir13}.~Detection limits for the Herschel PACS (Photodetecting Array Camera and Spectrometer) were flux ratios f = 10$^{-6}$ for $T({\rm dust})\!=\!50$\,K, and f = 10$^{-5}$ for $T({\rm dust})\!=\! 200$\,K.~Given these limits, it is interesting that nearly all of the FGK stars with detected debris disks are within $\sim\!20$ pc, and constitute $\sim$1/4 of their sample.~A-type stars with debris disks were detected to 40\,pc, and have cool dust temperatures ($\sim\!100$\,K) and small sizes ($<\!100$ AU), similar to the $\zeta$ Lep example above.~While Herschel observations are at 70, 100, and 160 microns, SED comparisons to the observations imply there would be very little mid-IR emission even at 20 microns in most of these systems. 

The HMP and other stars in this paper are much farther than those in the DUNES and DEBRIS surveys \citep[1 to 10\,kpc, e.g.][]{nor13}, and therefore are below the Herschel detection limits.~Thus, it remains a challenge to detect cool debris disks in these systems with the exquisite detail of other Herschel survey targets, even if debris disks are present.~Nevertheless, the question is whether formation of these debris disks could affect the stellar chemistry.~We address this issue in the next section.

\subsection {Debris Disks and Stellar Chemistry} 

The strongest evidence for the chemical anomalies that accompany dust-gas winnowing comes from the detailed abundance analysis of the atmosphere of a star for which the composition can be inferred {\it in the absence of dust-gas winnowing.} This inference is possible for Lambda Bootis stars, post-AGB stars and RV Tauri variables because (i) a wide range of elemental abundances are available for these stars, particularly over a broad range in the condensation temperature, and (ii) the compositions of {\it normal} dwarfs and giants of comparable metallicity span a small range in element abundance ratios, i.e.~$\Delta$[X/Fe] at given metallicity is small and therefore the magnitude of an abundance anomaly can be directly inferred despite a lack of knowledge of the star's original or intrinsic composition.~Unfortunately, this satisfactory state hardly applies to the HMP stars because (i) elemental abundances are available for only a few elements, (ii) key elements such as S and Zn provide few or no lines in their optical spectra, and most importantly (iii) at extremely low metallicities there are strong star-to-star differences which confuses our ability to identify chemical peculiarities from intrinsic properties.

If dust-gas winnowing is influencing the atmospheric composition of the HMP stars, then this mechanism not only predicts the presence of a dusty debris disk, circumstellar shell, or stellar wind, but it also requires (i) a mechanism such as differential radiation pressure between dust grains and gas enabling the atmosphere to accrete gas preferentially over dust, and (ii) a shallow atmosphere that is effectively unmixed with the convective envelope below.  Therefore, a theoretical estimate of the mass of the dust-free envelope mixed with the normal stellar atmosphere is a key factor; e.g.~$\sim\!100\times$ more dust-free gas is needed to reduce the [Fe/H] metallicity by 2~dex.~However, operation of the dust-gas winnowing mechanism is also governed by certain timescales.~For example, if the dust reservoir is exhausted or eliminated, a dust-gas winnowing chemical signature could remain without detectable dusty debris.~Alternatively, a debris disk or circumstellar dust envelope could last longer than chemical anomalies if rapid mixing with the deeper convective envelope erases the chemical signatures in the atmosphere.

The case of the Lambda Bootis stars highlights this latter scenario.~A study of 24 dusty A-stars by \cite{ack04} found that only one showed a clear chemically peculiar abundance pattern.~Thus, the presence of a debris disk indicated by a mid-IR excess is not a sufficient condition for operation of dust-gas winnowing affecting the chemical abundances in a stellar atmosphere.~Theoretical studies have suggested that the effects of dust-gas winnowing may last only $10^6$ years \citep{gas08,tur93}.

At the heart of the phenomenon is the presence of a thin radiative outer zone, where the composition can be influenced greatly by the accretion of a modest amount of dust-free gas.~The effective temperature of the main sequence Lambda Bootis stars ensures a thin outer radiative zone; post-AGB stars within the instability strip, such as the RV Tauri variables, as well as those bluewards of that strip, also have thin envelopes.~Detailed examination of the RV Tauri variables \citep[e.g.][]{gir05} have found (i) the dust-gas winnowing is only observed in stars with [Fe/H] $\le -1$, presumably because radiation pressure on the few grains of lower metallicity stars are unable to sustain the separation of dust from gas, and (ii) the phenomenon is only observed in stars hotter than 5000\,K, presumably because the atmospheres of cooler stars have deep convective envelope that would erase the signature.~As demonstrated in Section~2.2, for stars that do undergo dust-gas winnowing, the strength of the infrared excess is uncorrelated with the chemically peculiar abundance pattern, as with the Lambda Bootis stars, showing that a dusty reservoir is not a sufficient condition.

Cumulative indirect evidence for dust-gas winnowing affecting the HMP stars is therefore not encouraging.~As this paper shows, lack of a mid-IR excess appears to be common. The intrinsic metallicity, even if taken from C, N, and O abundances, is at least 1 dex below the limit suggested from RV Tauri variables (and the Lambda Bootis stars).~Many of the HMP stars are giants, with convective envelopes that would quickly erase most changes to the surface chemistry; the convective envelopes would also hamper the establishment of abundance anomalies through dust-gas winnowing in the first place.~Even if a young HMP star experiences an episode of dust-gas winnowing {\it early} in its life, signatures from this episode should be erased over the long lifetime of the star. 

Nevertheless, given the great diversity already exhibited by the five known HMP dwarf and giant stars, it may be that a star is found with a surface composition that is dominated not by ejecta from a first generation of stars but by the local operation of dust-gas winnowing.~The question is whether we have yet found that star, which we address in the next section. 

Finally, we note that the [Mg/Fe] ratio is also an interesting abundance ratio for this discussion.~Since Mg and Fe have very similar condensation temperatures, then HMP stars like \SMSSx\, with very high [Mg/Fe] ratios \citep{kel14} cannot be dominated by the effects of dust-gas winnowing.~In the case of \SMSSx, the high ratios of [C/Mg] and [Mg/Ca] further favour formation from gas enriched by faint supernovae in the early interstellar medium due to differentiation in the progenitor core and a mass cut for fall back.

\subsection{An IR Excess Associated with HE0107-5240?}

In this paper, we show a marginal (2 $\sigma$) detection of a mid-IR excess near 10 microns associated with the star HE0107-5240.~If this excess is real and due to thermal emission from a dusty circumstellar disk or shell, then a substantial amount of dust would need to be present given the distance to the star of 9620\,pc \citep{nor13}; this is $\sim\!500\times$ further than the stars in the Herschel telescope debris surveys ($\sim\!20$\,pc, described above).~The cause of the mid-IR excess (if real) is not clear; possibilities include (i) dust-gas winnowing, (ii) unassociated emission from the foreground interstellar medium, or perhaps (iii) effects in a binary system. 

Dust-gas winnowing does predicts an IR excess associated with stars where the mechanism operates (a prediction which motivated this paper), however this is clearly an uncommon occurrence in the HMP stars and cannot account for the chemical abundance anomalies of the entire class.~If dust-gas winnowing has occurred in HE0107-5240, then it must have been a local and recent phenomenon \citep[e.g.~a shock to the circumstellar envelope during passage through a dense portion of the interstellar medium as proposed by][]{gas08}.~Even then it is difficult to understand how dust-gas winnowing would work in a very metal poor star with [Fe/H]~$<\!-1$.

Unassociated emission from the foreground interstellar medium could cause broad non-thermal emission bands between 3 and 20 microns.~\cite{kwo11} have studied broad emission features seen in the infrared spectra (from ISO, the Infrared Space Observatory) of planetary nebulae, proto-planetary nebulae, photodissociation regions, 
and novae.~A number of aromatic and aliphatic organic (C-based) compounds\footnote{It is worth noting that these are not polycyclic aromatic hydrocarbon (PAH) molecules which require UV photons to excite them and yet often there are no UV sources available when these broad emission features are observed \citep[e.g.,][]{uch00}.} can cause these broad emission bands, with two particularly strong bands centered near 8 and 12 microns.~Again, given the distance to HE0107-5240, then it seems unlikely that these features would be associated with that stellar environment, although its C-rich chemistry would provide the basis for the formation of these complex organic molecules.

Finally, we notice that the dust-gas winnowing mechanism tends to occur in binary systems; the post-AGB stars are all spectroscopic binaries, and most (if not all) of the RV Tauri variables are binaries \citep{gir05}; however, it is unclear how binarity plays a role in the Lambda Bootis phenomenon \citep[if at all, e.g.][]{gri12}.~A binary system can provide the stability necessary to form a long-lived debris disk\footnote{Debris disks are defined as any small
objects (2000 km in diameter to submicrometer-sized dust) associated with a star, which includes the Solar System \citep{wya08}.~The stars targeted in the Herschel satellite DEBRIS survey are young \citep[0 to 1 Gyr;][]{boo13}, while the FGK stars in the DUNES survey have a wide range in ages \citep[0.1 to 13.6 Gyr;][]{eir13}.}~In an examination of the chemical properties of the HMP stars HE0107-5240 and HE1327-2326, \cite{tum07b} proposed that these stars could form from gas enriched by a primordial faint supernova, from gas enriched by a core-collapse supernova and C-rich gas ejected from the winds of massive stars, and/or as the low-mass secondaries in hyper metal poor binary systems, where they acquire their light-element enhancements from mass transfer when the companion passes through the AGB phase.~Following on the latter suggestion, the very low abundance of Li (and high Sr) in HE1327-2326 does favor the binary mass-transfer hypothesis; i.e.~in a typical AGB star, Li is destroyed through convective mixing, C is enhanced after the third dredge up, and Sr is enhanced through the s-process during the thermal pulsing stage, thus a low mass star that is C-rich, Li-poor, and Sr-enhanced is consistent with one that has been polluted by AGB mass transfer in a binary system.~The Li abundance in HE1327-2326 is well below the Spite plateau and primordial abundance from Big Bang nucleosynthesis, which is surprising for a dwarf or sub-giant star with \teff~$\approx\!6000$\,K \citep{fre05}.~The low Li abundance in HE0107-5240 is more complicated to interpret since it may have depleted its own Li during its evolution up the red giant branch.~Therefore, if the mid-IR excess in HE0107-5240 is real, it may indicate a debris disk formed in a binary system, and could indicate that some C-rich HMP stars are the predicted low mass, low metallicity stars that form in binaries. This also implies that the C-normal metal-poor star \SDSSx\ may be different if it has formed in isolation.

\section{Conclusions}
We present the mid-IR fluxes of the HMP and other metal-poor stars compared with predictions from standard stellar model atmospheres to test for the presence of a dusty circumstellar environment.~No mid-IR excesses are found for four out of five stars, ruling out may types of circumstellar disks.~We discuss the mid-IR excesses that are found associated with chemically peculiar stars in terms of dust-gas winnowing, and conclude that there is no evidence that dust-gas winnowing dominates the chemical abundance patterns of the HMP stars.~Only HE0107-5240 shows a marginal (2 $\sigma$) excess near 10 microns.~If the mid-IR excess is real and associated with HE0107-5240 (and not interstellar emission along the line of sight), we propose this may indicate a local and recent phenomenon leading to some dust-gas winnowing or the presence of a binary system.

\acknowledgments
We thank Prof.~John E. Norris for helpful comments on this manuscript, and Dr.~Brenda Matthews for many helpful discussions on debris disks.~KAV thanks NSERC for support through the Discovery Grant program.~THP acknowledges support by CONICYT through FONDECYT/Regular Project No.~1121005, and BASAL Center for Astrophysics and Associated Technologies (PFB-06), Conicyt, Chile.~DLL thanks the Robert A. Welch Foundation of Houston, Texas (grant F-634) for their support.~This publication makes use of the data products from the Wide-field Infrared Survey Explorer, which is a joint project of the University of California, Los Angeles, and the Jet Propulsion Laboratory/California Institute of Technology, funded by the National Aeronautics and Space Administration.~This publication makes use of the data products from the Two Micron All Sky Survey, which is a joint project of the University of Massachusetts and the Infrared Processing and Analysis Center/California Institute of Technology, funded by the National Aeronautics and Space Administration and the National Science Foundation.

{\it Facilities:} \facility{Gemini (T-ReCS), \facility{WISE}, \facility{2MASS}}.




\appendix
\vspace{0.25cm}
\section{Gemini Observations and Data Reductions}

Before the WISE All Sky Survey data release, observations at 10 microns were carried out for two stars, HE0107-5240 and HE0557-4840, with T-ReCS \citep{tel98} at the 8-m Gemini-South Telescope (August and September of 2009 as part of GS-2009B-Q88).~Here, we describe those observations and the data reduction steps 
carried out to obtain $N$-band magnitude limits.
 
Imaging was carried out at 10.2 microns ($\Delta\lambda$ = 1 micron), the $N$-band, to search for an infrared excess around the target stars and test for the presence of dust.~The pixel scale was 0.09\arcsec\ per pixel. To cancel out the background radiation, the secondary mirror chopping and the telescope nodding method were used.~We used four standard stars (HD\,720, HD\,14641, HD\,26967, and HD\,40808) from \cite{coh99} as flux calibrators and reference point spread functions. One standard star was observed before a science target, followed by one after, to ensure they were taken under similar observing conditions and at similar air masses.
 
The Gemini T-ReCS integration time calculator (ITC) was used to estimate the exposure times;~given the target $K$-band magnitudes, atmospheric temperatures, and an ATLAS9 model atmosphere, then 30-minute exposures were determined for the science targets.~The standard Gemini reduction package {\sc Mid-IR} was used within IRAF\footnote{Mid-IR data reduction web pages at: "http://www.gemini.edu/sciops/instruments/midir-resources/data-reduction". IRAF is distributed by the National Optical Astronomy Observatories, which are operated by the Association of Universities for Research in Astronomy, Inc., under cooperative agreement with the National Science Foundation.}. Simple averaging of the stacked images was used in the {\sc Mistack} task to co-add the chopped and nodded frames.

For HE0107-5240, the standard stars HD\,720 and HD\,14641 were observed before and after the science observation.  Each standard star was measured using the IRAF task {\sc Imexamine} to determine the flux and FWHM as listed in Table~\ref{tab:obs} within an aperture radius of 15 pixels, a buffer of 10 pixels and a sky annulus width of 25 pixels.~Using the known flux density at 10.2 microns of the standard stars (interpolated from the mid-IR fluxes on the Gemini webpages, also \cite{coh99};~listed in Table~A1), a conversion factor of $2.769\times 10^{-6}$ Jy/ADU was determined.~The flux of standard star HD\,14641 was measured using the same aperture.~The conversion factor determined for HD720 was then applied to HD\,14641, which yields a flux density of 6.59 Jy, in excellent agreement with the known flux of 6.59 Jy (see Table A1).

\begin{center}
\begin{deluxetable*}{lrcccc}
\footnotesize
\tablecaption{Journal of Observation and Flux Densities\label{tab:obs}}  
\tablewidth{0pt}
\tablehead{
\colhead{Target} & \colhead {Spectral} & \colhead{Integration} &  \colhead{FWHM} & \colhead{Flux} & \colhead{Flux Density$^{a}$ }  \\
\colhead{} & \colhead{Type}  & \colhead{Time (s)} & \colhead{(pix)} & \colhead{(ADU)} & \colhead{ Jy } 
} 
\startdata
HD720 & K2 III & 	29 &  5.05  & $1.694\times10^6$     & 4.69  \\
HE0107-5240 & \nodata & 1853$^{b}$ &  5.04$^{c}$  & $\le 7.719\times10^3$ &  $\le 0.33\times 10^{-3}$ \\
HD14641 & K5 III & 	29 &  5.02  & $2.381\times10^6$     & 6.59  \\
HD26967 & K1 III & 	29 &  3.67  & $4.381\times10^6$     & 11.91 \\
HE0557-4840 & \nodata & 1824 &  5.04{$^c$}  & $\le 4.597\times10^3$ &  $\le 0.20\times 10^{-3}$ \\
HD40808 & K0 III & 	29 &  3.55  & $4.012\times10^6$	    & 10.67 \\\vspace{-0.2cm}
\enddata
\tablecomments{{$^a$}The flux densities for the standard stars are from the Gemini webpage and Cohen et al. (1992; see http://www.gemini.edu/sciops/instruments/mir/Cohen\_list.html). {$^b$} A 3 $\sigma$ limit was determined using $3\times$FWHM, also the larger aperture FWHM for HE0557-4840 was adopted based on the observations of HE0107-5240. {$^c$} An additional observation of HE0107-5240 was made on 2009-08-30, but interrupted due to cirrus cloud cover and was not used in this analysis.}
\end{deluxetable*}
\end{center}

To calculate a 3 $\sigma$ upper limit for HE0107-5240, we considered an aperture that was three times larger than that measured for the standard stars (i.e.~3 times their average FWHM, or $3\times 5.035$ pixels) centered around the expected position of HE0107-5240.~The mean value of the counts in this aperture was 10.75 ADU {\it per pixel}, with a standard deviation of 13.25 ADU.~The mean value was multiplied by the number of pixels and converted to flux by the factor found above for the standard stars.~This mean value was then scaled by the exposure time ratio (28.9585 sec/1853.34 sec).~The flux density upper limit for HE0107-5240 is then 0.33 mJy.~This was converted to a magnitude by using the known zero magnitude flux density of 38.5 Jy from the Gemini Flux Density/Magnitude conversion tool\footnote{http://www.gemini.edu/?q=node/11119}, giving an apparent N magnitude upper limit of, 

\begin{equation}
 {\rm N}_{\rm HE0107-5240} \le 12.66 \ \ {\rm mag}
\end{equation}

The same procedure was followed for HE0557-4840, using HD\,26967 and HD\,40808 as standard stars (see Table~A1).~The mean value per pixel in the $N$-band image of HE0557-4840 was 6.40 ADU.~Small differences between the Jy/ADU conversion factor determinations for the standard stars meant that we applied $2.690\times10^{-6}$ Jy/ADU to the HE0557-4840 data.  Also, the sharp image profiles of the standard stars may have been due to variable seeing during the observations, therefore we adopt the more conservative FWHM = 5.04 pixels determined from HE0107-5240 for the calculation of the flux density limit for HE0557-4840.~As above, this was converted to an $N$-band magnitude of, 

\begin{equation}
 {\rm N}_{\rm HE0557-4840} \le 13.21 \ \ {\rm mag}
\end{equation}

The corresponding N-band flux densities are illustrated in Figure~1, along with the optical and near-IR values.

Due to very low background levels for the standard star exposures (typically 50 ADU), the uncertainties in their flux density values are dominated by the calibration of the spectral irradiance values estimated as $\simeq 2\%$ by \cite{coh92}, combined with the mid-IR filter set estimated to be on the order of $\simeq$ 2\% by Gemini's online resource page.~No air mass corrections were applied since the standard stars and science targets were all observed at approximately the same air mass. No color-temperature correction terms were applied either since  the standard stars have similar temperatures as the science targets, in the range of 3800-5000\,K (spectral types of the standard stars are in Table~A1).~We do not consider differential reddening effects either as all of these stars are bright and nearby.

\end{document}